\documentstyle[emulateapj,onecolfloat,psfig]{article}   

\begin{document}
\twocolumn[      

\title{Scintillation Caustics in Planetary Occultation Light Curves}

\author{Asantha R. Cooray\altaffilmark{1}, J. L. Elliot\altaffilmark{2}}
\affil{$^1$Theoretical Astrophysics, California Institute of Technology, Pasadena, CA 91125\\
$^2$Department of Earth, Atmospheric \& Planetary Sciences
and Department of Physics,
Massachusetts Institute of Technology, Cambridge, MA 02139.}


\begin{abstract}
We revisit the GSC5249-01240 light curve obtained during its occultation by
Saturn's North polar region. 
In addition to refractive scintillations, the power spectrum of intensity fluctuations shows an enhancement of 
power between refractive and diffractive regimes. We identify this excess power as due to
high amplitude spikes in the light curve and suggest that these spikes are due to caustics
associated with ray crossing situations. The flux variation in individual spikes follows the expected caustic behavior,
including diffraction fringes which we have observed for the first time in a planetary occultation light curve.
The presence of caustics in scintillation light curves require an inner scale cut off to the power spectrum of 
underlying density fluctuations associated with turbulence. Another possibility is the presence of gravity waves
in the atmosphere. While occultation light curves previously showed the existence of
refractive scintillations, a combination of small projected stellar size
and a low relative velocity during the event have allowed us to identify caustics in this occultation.
This has led us to re-examine previous data sets, in which we have also found likely examples of caustics.
\end{abstract}

\keywords{occultations---planets: atmospheres---planets: Saturn}
]

\section{Introduction}
The occultation of background stellar sources by foreground solar system objects
provides one of the well utilized probes of distant 
planetary atmospheres (for a review, see, Elliot \& Olkin 1995). 
The light curve produced during such events allow 
one to extract detailed information related to the thermal
structure of the atmosphere 
with spatial resolutions at the distance of the planet of order of a few
kilometers.  In recent years, increasingly sensitive instrumentation 
has allowed occultation observations with spatial accuracies not limited by instrumental effects, but rather by physical parameters associated with the occultation geometry---such as the distance, the relative velocity, and 
the projected size of the occulted star at the planet.

In Cooray et al. (1998), we 
presented results from an occultation by Saturn's North-polar region of the 
star GSC5249-01240 on November 20, 1995, 
near the time of the solar-ring plane crossing of Saturn.
 This occultation was predicted by Bosh \& McDonald
(1992) to occur with the slowest relative velocity, 0.59 km sec$^{-1}$,
of their 203 predicted stellar occultations by Saturn between
1990 and 2000. The small projected diameter of the background star, $\sim$ 0.1 km, 
allowed sub-kilometer spatial resolution of the Saturn's atmosphere.

The light curve of GSC5249-01240 obtained during 
the occultation at NASA's Infrared Telescope Facility (IRTF)
in Mauna Kea, Hawaii showed an excessive 
number of high amplitude spikes in the light curve. 
In Cooray et al. (1998),
we tentatively identified these 
large amplitude variations, of over 100\% in some occasions, 
as due to diffractive scintillations. This suggestion
was primarily based on the fact that
the projected size of the star at the distance of Saturn, $\sim$ 0.1 km, was much smaller than the
Fresnel scale associated with these observations, $\sim$ 0.7 km. 
As the light curve probes deeper in to the atmosphere, 
 scintillation fluctuations observed in such cases
 are well described through a combination of weak and strong  scintillation theories
based on refraction alone (e.g., Narayan \& Hubbard 1988). 
In the present occultation, due to the smaller size of the star when compared to the 
Fresnel scale,  we entered a regime for the first time where 
potentially new and interesting signatures may be observed.

We recommend the reader to Cooray et al. (1998) for observational details. There, we also
present an analysis based on the temperature structure of Saturn's North polar region.
Here, we  study the light curve under 
the context of scintillation theories based on both
diffraction and refraction and discuss the power spectrum of
intensity fluctuations as well as individual high amplitude spikes. 
We show that the light curve contains scintillation caustics
involving multiply imaged events; the presence of caustics in our light curve
agrees with theoretical  predictions made previously 
by Goodman et al. (1987) with regards to caustics in
pulsar light curves due to interstellar  scintillation. Similarly, while not recognized as
such, the same caustics events have also been predicted by French \& Lovelace (1983)
due to ray crossing situations associated with density fluctuations generated by
gravity waves. Our light curve probably provides the best evidence for caustics --- whatever their cause.

\section{Scintillation Caustics}

The application of scintillations to stellar occultation
light curves, either due to turbulence or coherent gravity waves,
 has been widely discussed in the literature
(e.g., Young 1976; French \& Lovelace 1983; Hubbard \& Narayan 1998).
In general, scintillation processes are expected from fluctuations in the underlying density field,
or more appropriately the refractivity in the case of stellar occultation 
light curves, with a power-law spectrum of the form:
\begin{eqnarray}
Q(\vec{k})&=& Q_0 \left[k^2 +k_{\rm out}^2\right]^{-\beta/2} \exp\left(-\frac{k^2}{2k_c^2}\right) \nonumber \\
 &\approx& k^{-\beta}\; \quad \quad \quad \quad  k_{\rm out} < k < k_c\, ,
\end{eqnarray}
where $k$ is the spatial wave-number and 
$k_{\rm out}$ and $k_c$ are outer and inner cut-off scales of the fluctuation power spectrum. 
In the case of an exponential atmosphere, the outer cut-off  scale is related to the atmospheric
scale height, $H$ (French \& Lovelace 1983).

Between the two cut-off scales, 
the usual assumption of a power law in density fluctuations leads to two separate and well 
known regimes of scintillations involving refraction and diffraction. 
If the power density spectrum has an inner scale cut off between the
refractive and diffractive regimes, Goodman et al. (1987) showed that there is excess power
at the same intermediate regime due to caustic events. In the case of stellar occultations, 
these caustics can be associated with ray-crossing situations, either due to turbulence or
gravity waves (French \& Lovelace 1983).
The flux power spectrum in the presence of caustics can be broken to three
regimes, refractive, intermediate, and diffractive with
\begin{eqnarray}
&&k^2P(k) = \nonumber \\
&& \left\{ \begin{array}{ll} 
Q_0 r_F^\alpha (kr_F)^{4-\alpha}  \; \exp 
\left[-\frac{1}{2}\left(\frac{k}{k_{\rm ref}}\right)^2 \right]\; \quad \quad k < k_{\rm cref} \, \\
\frac{\pi}{\sqrt{2} \alpha(2-\alpha) |A|} (r_c/r_F)^{4-\alpha} \,  \quad \quad \quad \quad k_{\rm cref} < k < k_{\rm cdif}\; \\  
\frac{\pi}{\alpha|A|} (r_c/r_F)^{2-\alpha} (kr_F)^{2}  \exp\left[-\frac{1}{2}\left(\frac{k}{k_{\rm dif}}\right)^2\right]\;  \quad k > k_{\rm cdif} \, , 
\end{array} \right.
\label{eqn:power}
\end{eqnarray}
respectively, where  $\alpha=\beta-2$. Following Goodman \& Narayan (1985), we write
\begin{equation}
A = \frac{\Gamma[(4-\alpha)/2]}{2^\alpha \pi \alpha (\alpha-2) \Gamma[(\alpha+2)/2]} Q_0 r_F^{\alpha} \, .
\end{equation}

The various length scales involved in Eq.~2 are (1) 
the Fresnel scale, $r_F = (\lambda d/2\pi)^{1/2}$,
 when wavelength of observations is $\lambda$ and
distance to the scattering screen is $d$, (2) 
the refractive scale, $r_{\rm ref} = \theta d$,
when the scatter-broadened image size is $\theta$ (Eq. 2.10 of Narayan \& Hubbard 1988), 
(3) the diffractive scale, $r_{\rm dif}=r_F^2/r_{\rm ref}$, that determines the fringe separation,  
(4) the inner cut off scale of density fluctuations, $r_c=2\pi/k_c$, (5)
the cut off scale of refractive scintillations, $r_{\rm cref}=r_{\rm ref}$ if $\beta <4$ or
 $r_{\rm cref}=r_{\rm ref}(r_c/r_{\rm ref})^{(\beta-4)/2}$, if $\beta >4$, and, finally, (6)
the cut off scale associated with  diffractive scintillations, $r_{\rm cdif}=r_F^2/r_c$. These length scales are
inter-related such that $r_{\rm ref}r_{\rm dif}=r_c r_{\rm cdif}=r_F^2$. 
Here, length scales, $r$, are related to wave numbers via $k=2\pi/r$.  In the case of stellar occultations, 
note that the Fresnel scale decreases as one probes deep in to the atmosphere.

In the presence of an inner-scale cutoff, which lies between the refractive and diffractive scales,
$k_{\rm cref} < k_c < k_{\rm cdif}$, one expects the power spectrum of
intensity fluctuations, $P(k)$, to scale as $k^{-2}$ in the same intermediate regime 
between diffractive and refractive scintillations. 
The presence of refractive scintillations requires  that the outer cut off scale
satisfy the condition $r_{\rm out} \gg r_{\rm ref}$ (Goodman et al. 1987).

\begin{figure}[t]
\centerline{\psfig{file=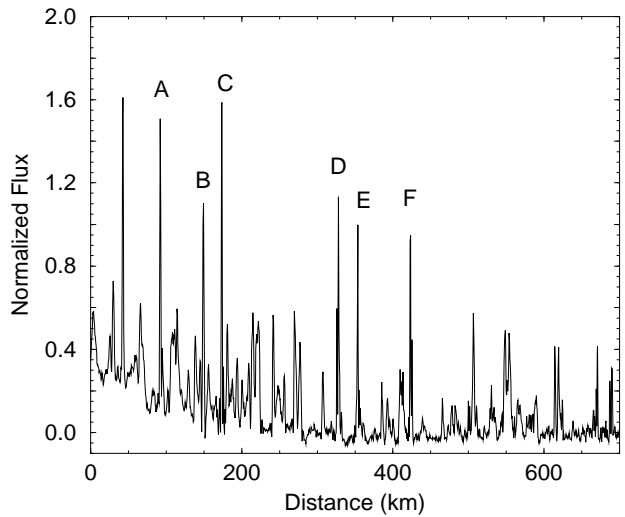,width=3.2in,angle=-90}}
\caption{The GSC5249-01240 light curve during the occultation by Saturn's North polar region
and discussed in Cooray et al. (1998). The existence of high amplitude short duration spikes are
clearly visible. Note that these appear as pairs, which is an indication for caustics. 
The labeled spikes are shown in Fig. 3.}
\label{fig:lc}
\end{figure}

In Fig.~1, we show the light curve of GSC5249-01240 during the occultation by Saturn. 
The existence of high amplitude spikes are clearly visible.
The relevant parameters for the GSC5249-01240 event are: 
relative velocity, $v_{\perp} =0.586$ km sec$^{-1}$, 
distance to Saturn, $d=9.194$ AU, wavelength, 
$\lambda=2.28$ $\mu$m, and integration time, $t_{\rm int} = 0.947$ sec.
The Fresnel scale is $\sim$ 0.71 km while the resolution
of the light curve is $\sim$ 0.55 km. Based on the spectrum of the star, the projected radius of the
star at the distance of Saturn was determined to be of order 0.1 km.
For the rest of the discussion, we renormalize the light curve based on model fits presented in Cooray et al. (1998) and remove the variation in mean flux and only consider fluctuations
with respect to this mean. 

\begin{figure}[t]
\centerline{\psfig{file=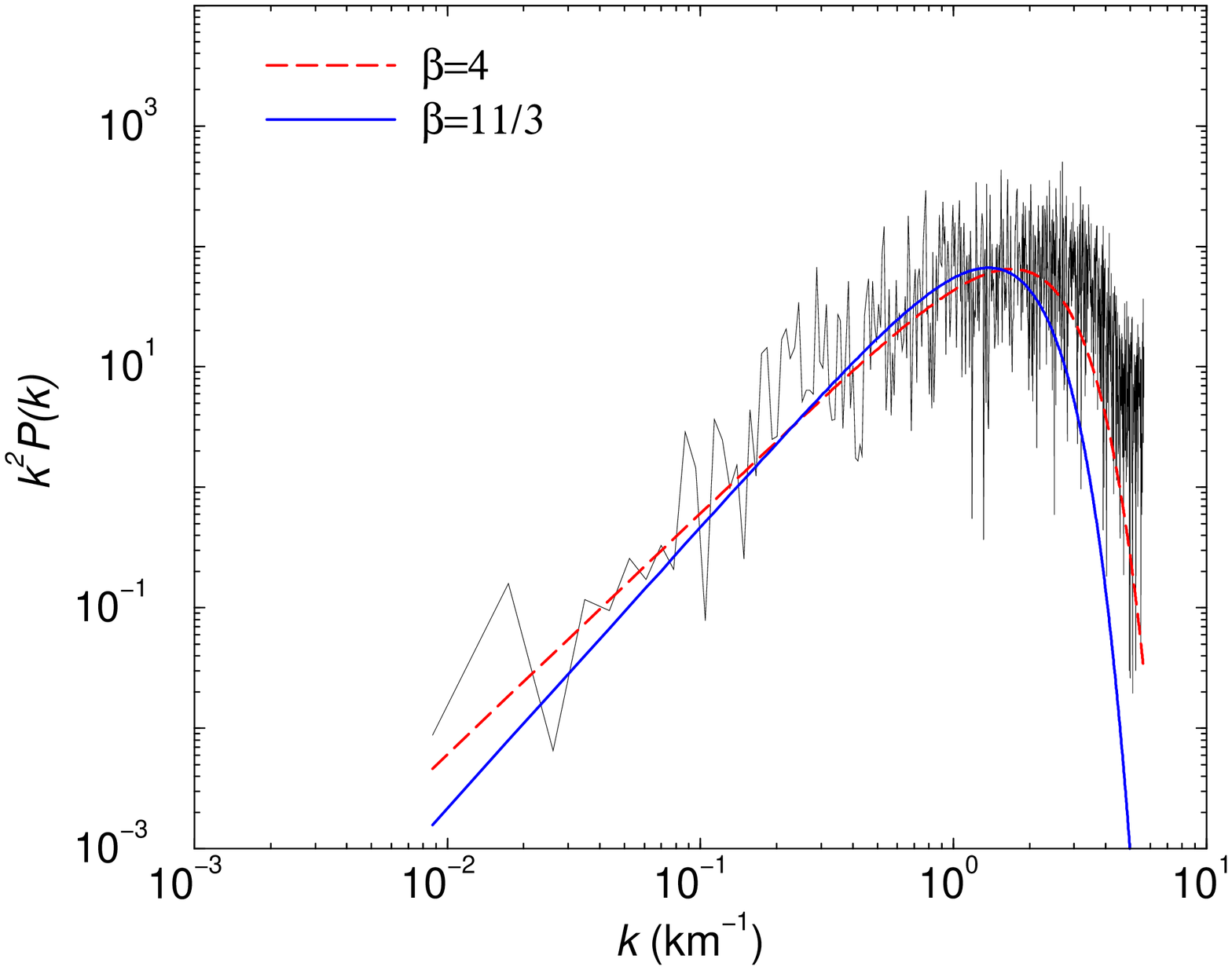,width=3.0in,angle=-90}}
\centerline{\psfig{file=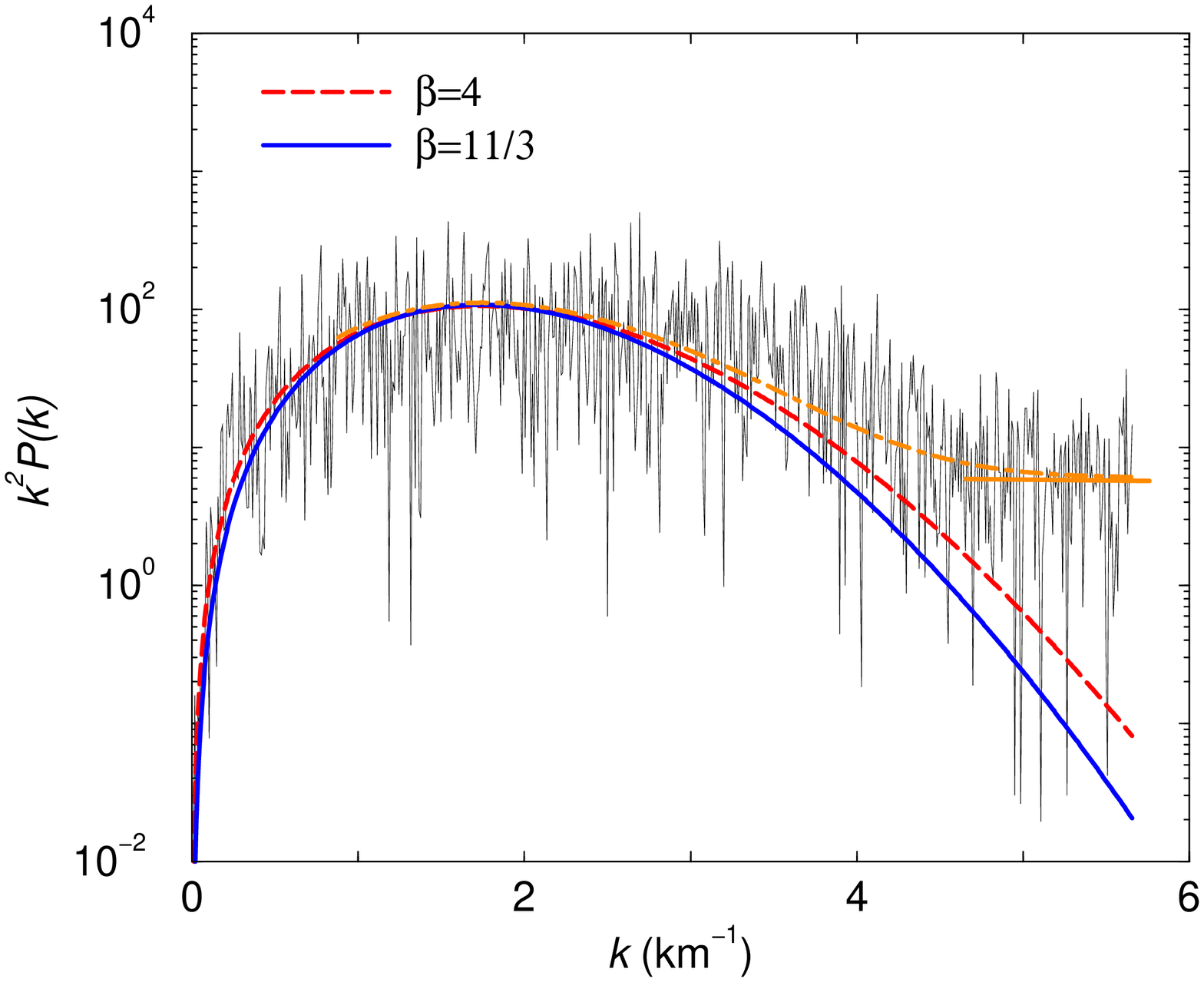,width=3.0in,angle=-90}}
\caption{The power spectrum of intensity fluctuations. 
The solid and dashed curves are 
based on fits to the power spectrum with wavenumbers less than 3
km$^{-1}$ and using Eq.~2 with assumed power-laws for the
underlying density-field fluctuation power spectrum. 
The peak, at wavenumbers near 1 km$^{-1}$, 
is associated with refractive scintillations, while an additional peak, which we have not probed,
is expected at high wavenumbers due to diffractive scintillations.
In the bottom panel, we concentrate on the 
high wave-number region where, to the right, we see an indication for the flattening of power at wavenumbers
above 4 km$^{-1}$.  Such a flattening is expected when caustics are present in the light curve in addition to simple
refraction (Goodman et al. 1987). }
\label{fig:power}
\end{figure}

The power spectrum of intensity fluctuations, $P(k)$, is presented in Fig.~2 where 
we plot $k^2P(k)$, which represents equal weighing in logarithmic 
wavenumber intervals. In general, this
power spectrum is expected 
to show two distinct peaks associated with refractive and diffractive
scintillations (see, for e.g., Fig.~3 of Goodman et al. 1987). 
The data clearly show the existence of the first peak associated with
refractive scintillations. In the presence of an inner-scale 
cut off to the underlying density field
fluctuations, we expect excess power beyond the refractive peak and a flattening of power at large wave numbers. 
We see an evidence for such an increase in power and flattening as shown in the
bottom plot of Fig.~2.
Here, we have focused on only the large wave number regime that is of
interest. 

The two solid and dashed curves show fits to the
power spectrum under the assumption of $\beta=11/3$, 
a Kolmogorov type spectrum, and $\beta=4$ in Eq.~2.
With the slope fixed, the model fitting involves two 
unknown free parameters related to $Q_0$ and $k_{\rm ref}$.
In order not to bias model fitting with respect to the excess, we make
use of the power spectrum only out to wave numbers less than 3 km$^{-1}$;
While we can obtain a fit to the whole power spectrum,
we, however, cannot explain the apparent flattening of power
beyond the turnover at wave numbers greater than $\sim$ 4.5 km$^{-1}$. 
Note that the
flattening cannot be ascribed to simply noise in the power spectrum; white
noise fluctuations will lead to scale independent power with $P(k) \sim$ 
constant such that in a $k^2P(k)$ plot, one would see a sharp rise.

Using fits to the power spectrum based on theoretical expectations, following 
Eq.~2, we determined that for a $\beta=11/3$ power law, $r_{\rm ref}=r_{\rm cref} = 5.7 \pm 1.2$ km, and 
$Q_0 = 380 \pm 130$, at the one-sigma confidence level. For a $\beta=4$ power law, these numbers are
$5.2 \pm 1.1$ km and  $370 \pm  140$ respectively. The errors in fitted parameters are estimated by
allowing the normalized light curve to vary randomly up to $\pm$ 5\%; this accounts, rather
conservatively,  relative photometric errors and any potential uncertainties associated with the
light curve. Our discussion is not subject to absolute 
calibration uncertainties, as we are only interested in relative changes.
The associated  diffraction scale is  $r_{\rm dif} \sim 0.09$ km or $k_{\rm dif}\sim 70$ km$^{-1}$;
we do not probe such small length scales, suggesting that we
have not observed regular diffractive scintillations in the present data.

We can determine the inner cut-off scale by comparing
the amplitude of flattening to that of the refractive peak in the power
spectrum. For the $\beta=11/3$ and 4 power-laws, we determine this cut off to
be at  $r_c \sim 4.0 \pm 1.5$ km and 4.4 $\pm 1.6$ km, respectively.  The
caustic regime extends from the cut off scale associated with
refraction, $r_{\rm cref} \sim$ 6 km, to the cut off scale associated with
diffraction, $r_{\rm cdif} \sim 0.12$ km.  Following Goodman et al. (1987), 
the expected flux variations associated with caustic events are given by
$(\Delta F/F)^2 \sim 2 (r_c/r_{\rm ref})^2 \ln (r_c r_{\rm ref}/r_F^2)$.
Based on derived values, we find $(\Delta F/F) \sim 1.9$ which is 
consistent, though some what higher than, the observed amplitudes of 
spikes in the light curve. 

The presence of caustics can also be verified by inspecting the intensity
profile of individual events. In general,
individual caustics are expected to show 
the $1/\sqrt{d}$ behavior in the flux variation with distance, $d$,
as the source moves away from the critical line. 
In the case of stellar occultation light curves, such
a distance dependence is expected in the observer plane
such that the intensity varies as $I(d) = I_0 +\Theta(d) a_0 d^{-1/2}$, 
where $\Theta(d)$ is the step function, $a_0$ is
the caustic strength, and $I_0 \sim 0$ is the background flux. 
This intensity dependence, at Saturn,  can be written using the fact that
the mapping between the source and observer plane involves 
$d = r_{\rm cref} (x/r_c)^2$  and $a_0 \sim r_c^2/r_{\rm cref}^{3/2}$ (Goodman et al. 1987),
where $x$ is measured distance in the scatterer plane.
This leads to an intensity decrease for high amplitude spikes as
$I(x) \sim A_0/x$, where $A_0  \sim r_c^3/r_{\rm cref}^2 \sim 1.9$.
In addition to this one-sided drop in flux, caustics are also expected to 
show the presence of diffraction with an  oscillatory pattern superimposed on the intensity variation. 

In Fig.~3, we summarize our model descriptions of individual high amplitude 
spikes. While the expected 
width of the caustic is a priori known, $\sim (r_{\rm ref} r_F^4/r_c^2)^{1/3}$ (Goodman et al. 1987), we have allowed for smoothing with 
a Gaussian profile, $\exp(-x^2/2\sigma^2)$, and treated the width as an unknown
parameter. Individual caustic curve model fits to spikes are 
determined by three parameters:
the caustic strength $A_0$,  smoothing width, and the oscillation spacing due to 
diffraction. From  fits to six spikes, we determine $A_0 = 1.8 \pm 0.4$,  consistent with the expectation value based on parameters derived from the power spectrum analysis.
The smoothing width, $\sigma$, is  determined to be of order $0.5 \pm 0.2$ km
and is  consistent with the expected width of order 0.45 km.
Note that the caustic width is higher 
than the size of the star, $\sim 0.1$ km, based on
its spectrum; While Goodman et al. (1997) considered the case of a background
point source, in our case with the finite size for the star, 
the final caustic width is likely to be smoothed by
the projected star size at the distance of Saturn.

Note that caustic spikes appear in pairs in the light curve; such a pairing
may lead one to conclude the possibility that the occultation, in fact, involves a binary instead 
of a
single star, as was observed in the occultation of $\beta$-Scorpii  by Jupiter (Elliot et al. 1976).
There is one observation that rules out this possibility; in our spikes, we see the alternating right- and left-handed 
behavior associated with the one-sided decrease in flux. This is
naturally associated with caustics.

Before discussing our main results, we note that
previous studies on refractive scintillations in occultation light curves
have shown evidence for anisotropy in underlying fluctuations 
such that in Eq.~1, $k^2 = \rho^2 k_x^2 + k_y^2$ with $\rho > 1$ 
(Narayan \& Hubbard 1988). While for 
the present discussion we ignored the presence of an anisotropic medium,
we plan to address this issue as part of a detailed analysis of several
light curves of the same occultation event obtained with the
Hubble Space Telescope. 

\begin{figure}[!h]
\centerline{\psfig{file=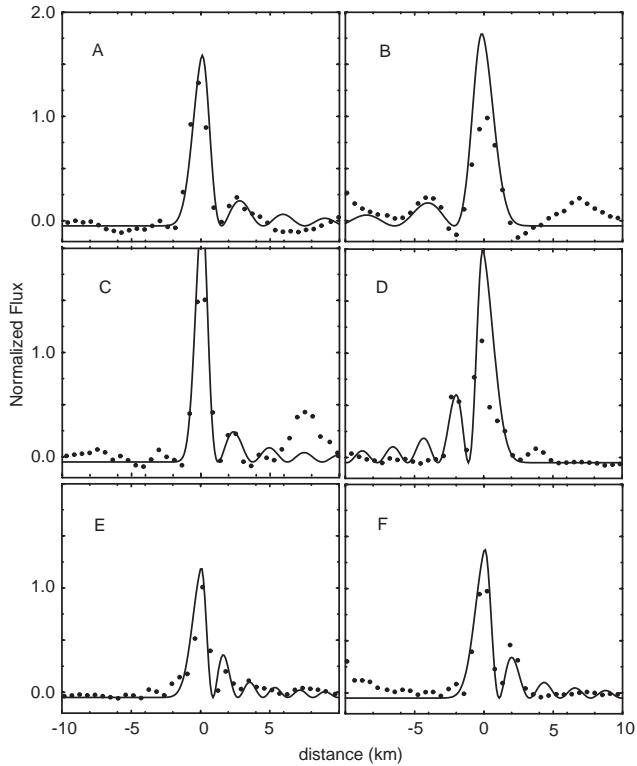,width=3.3in,angle=0}}
\caption{Individual spikes as labeled in Fig.~1. 
The model fits, shown as solid lines, are based on caustic descriptions
with an oscillation due to diffraction.
The pairs, A\&B and C\&D, show the expected right and left-handed one-sided
decrease in flux associated with caustics.}
\label{fig:spikes}
\end{figure}

\section{Discussion}
Among the two processes for strong scintillation, 
refractive scintillations have been considered
in the past as a source of fluctuations in the stellar occultation light
curves associated with giant planet atmospheres. 
The second, diffractive scintillations, was not expected to contribute
to intensity fluctuations given the required high resolution to see
its effects and the small stellar size required not to average out the 
diffractive pattern (e.g., Narayan \& Hubbard 1988).  
In the case of the occultation of GSC5249-01240 by Saturn,
the projected size of the star was smaller than the Fresnel scale and
the relative velocity during the event was of order 0.5 km sec$^{-1}$. This, in combination,
allowed sub-km spatial resolution.

The light curve contains distinct signatures of sharp and high amplitude spikes;
such a high number of them has not been observed previously.
While we explored the regime below refractive scintillations, 
the presence of diffractive scintillations require spatial resolutions less than $0.1$ km, below what we have probed.
The diffractive scintillations are  expected to produce flux variations of order 100\% 
at regular intervals. In our case, we see high amplitude spikes at irregular intervals 
with normalized fluxes well over unity suggesting that the observed  phenomenon is not typical.

The spikes, however, can be described through caustics whose
presence require an inner cut off to fluctuations in the underlying density power spectrum (Goodman et al. 1987). 
Such a cut off may also be responsible for ray crossing one needs for the formation of
caustics (French \& Lovelace 1983). Following the notation of French \& Lovelace (1983), the present light curve has
an intensity fluctuation strength, $A$, of 14 and a normalized star size of $d_\star=0.05r_F$.
The strength of intensity fluctuations suggests fractional fluctuations in the refractivity,
$\Delta \nu/\bar{\nu}$, of order $4 \times 10^{-3}$. If underlying fluctuations are due to gravitational waves,
under geometrical arguments, ray crossing is expected when vertical wavelengths are less than 
$2 \pi H/(\Delta \nu/\bar{\nu})^{-2/3} \sim H/6$ or 7 km since $H \sim 40$ km (Cooray et al. 1998) for this region 
of the atmosphere. This value is the constraint when only a
single wave is present in the atmosphere. The case  with a spectrum of
waves is analogous to that described by Goodman et al. (1987) involving turbulence.

The presence of caustics is aided by several observations that involve
the detection of excess power in the power spectrum of
intensity fluctuations and the expected 
flattening of the same power spectrum in the intermediate regime 
between refractive and diffractive scintillations.  The intensity variation in individual spikes is 
also well described through the caustic description with a modulation due to diffraction. The parameters
from the power spectrum are consistent with model descriptions to the spikes
aiding the suggestion for the presence of caustics. 
While we have not investigated in detail, we may have found occasional
examples of caustic spikes in previous occultation light curves, such as the  
``camel'' spike of Elliot \& Veverka (1976). These, however, do not contain the signature of diffraction fringes
as observed in the present occultation.

From our data, we determined the inner scale cut off of
the underlying density fluctuation power spectrum to be around 4.5 km.
The density fluctuations are generally considered to be caused by processes involving either 
turbulence or gravity waves (e.g. Elliot \& Veverka 1976; Young 1976; French \& Lovelace 1983;
Narayan \& Hubbard 1988).  
It is likely that the cut off may be associated with viscous dissipation of 
wave energy. In Cooray et al. (1998),  we found an
increase in temperature of $\sim$ 15 K between $\sim$ 10 and 1
$\mu$ bar of the region probed by this occultation. Using our cut off scale and parameters appropriate for
this occultation, we extend the calculation by Roques et al. (1994) on heating by wave decay to estimate an
energy transfer rate of $\sim 10^{-7}$ ergs cm$^{-3}$ s$^{-1}$. 

The detection of caustics and, for the first time, diffraction fringes in an occultation light curve  
suggest that we entered a regime which was previously unexplored and unexpected to be present in occultation light 
curves.  We believe that it was a combination of the small velocity,
high sampling rate, and the small projected size of the occulted  star that allowed us to probe
the intermediate regime between refractive and diffractive scintillations for the first time.
We suggest events with such favorable conditions be given additional priority for observation.

\smallskip
{\it Acknowledgments:} 
We thank R. Blandford, P. Goldreich, D. Holz, L. Koopmans, and R. Sari for useful discussions and helpful 
suggestions. We acknowledge
support from DOE DE-FG03-92-ER40701 and the Sherman Fairchild Foundation (at Caltech) and
NASA Grant NAG5-10444 and NSF AST-0073447 (at MIT).

\end{document}